\documentclass[sn-mathphys,usenatbib]{sn-jnl}

\usepackage{graphicx}	
\usepackage{amsmath}	
\usepackage{amssymb}	
\usepackage{comment}
\usepackage{natbib}

\newcommand{\kms}{\mbox{km~s$^{-1}$}}
\newcommand{\Msun}{\mbox{$\text{M}_{\odot}$}}
\newcommand{\alr}{\mbox{$^{26}$Al}}

\newcommand{\fer}{\mbox{$^{60}$Fe}}

\jyear{2022}%

\theoremstyle{thmstyleone}%
\theoremstyle{thmstyletwo}%

\theoremstyle{thmstylethree}%

\begin{document}

\title[Dynamics of planet-forming environments]{Dynamics of young stellar clusters as planet forming environments}


\author*[1,2]{\fnm{Megan} \sur{Reiter}}\email{Megan.Reiter@rice.edu}

\author[3]{\fnm{Richard} \sur{Parker}}\email{r.parker@sheffield.ac.uk}

\affil*[1]{\orgdiv{Department}, \orgname{ESO}, \orgaddress{\street{Karl-Schwarzchild Str. 2}, \city{Garching bei M\"{u}nchen}, \postcode{85748}, \state{Bavaria}, \country{Germany}}}

\affil[2]{\orgdiv{Department of Physics and Astronomy}, \orgname{Rice University}, \orgaddress{\street{6100 Main St.}, \city{Houston}, \postcode{77005-1827}, \state{Texas}, \country{USA}}}

\affil[3]{\orgdiv{Department of Physics and Astronomy}, \orgname{The University of Sheffield}, \orgaddress{\street{Hicks Building, Hounsfield Road}, \city{Sheffield}, \postcode{S3 7RH}, \state{South Yorkshire}, \country{UK}}}


\abstract{
Most stars and thus most planetary systems do not form in isolation. The larger star-forming environment affects protoplanetary disks in multiple ways: gravitational interactions with other stars 
truncate disks and alter the architectures of exoplanet systems; external irradiation from nearby high-mass stars truncates disks and shortens their lifetimes; and remaining gas and dust in the environment affects dynamical evolution (if removed by feedback processes) and provides some shielding for disks from external irradiation. The dynamical evolution of the region regulates when and how long various feedback mechanisms impact protoplanetary disks. Density is a key parameter that regulates the intensity and duration of UV irradiation and the frequency of dynamical encounters. The evolution of larger star-forming complexes may also play an important role by mixing populations. Observations suggest that clusters are not a single-age population but multiple populations with small age differences which may be key to resolving several timescale issues (i.e., proplyd lifetimes, enrichment). In this review, we consider stellar clusters as the ecosystems in which most stars and therefore most planets form. We review recent observational and theoretical results and highlight upcoming contributions from facilities expected to begin observations in the next five years. Looking further ahead, we argue that the next frontier is large-scale surveys of low-mass stars in more distant high-mass star-forming regions. The future of ecosystem studies is bright as faint low-mass stars in more distant high-mass star-forming regions will be routinely observable in the era of Extremely Large Telescopes (ELTs). 
}


\maketitle

\section{Introduction}\label{s:intro}

There is a long history in astronomy of considering the likely birthplace of the Solar System (e.g., \cite{adams2001,adams2004,hester2004,adams2006,looney2006,pfalzner2013,portegies-zwart2019,batygin2020,pfalzner2020}). 
Previous reviews (e.g., \cite{adams2010,pfalzner2015}) have considered the current features of the Solar System to infer the required birth conditions that lead to the present configuration. 
While the details are still debated, several key features are clear: 
the Solar System was likely born in a stellar cluster with at least one high-mass star that provided short-lived radioactive isotopes; and 
the cluster density was modest to prevent tidal effects disrupting the natal disk which would be inconsistent with the observed size of the Kuiper Belt.
Taken together, these considerations suggest that the Solar System formed in an Orion-like cluster (see discussion in \cite{adams2010}).

One of the early clues was the presence of short-lived radioactive isotopes discovered in Solar System meteorites, in particular $^{26}$Al and $^{60}$Fe. 
The short radioactive decay times for both of these elements (half-life of 0.717~Myr and 2.62~Myr for \alr\ and \fer, respectively) requires that the Solar System formed near a source of these elements in both time and space. 
As reviewed recently by \cite{lugaro2018}, the timescale is one of the most stringent limitations for how these elements were introduced into the Solar System. 
Studies of the pathway and likelihood of enrichment naturally led to the broader question of how often Earth-like planets (Solar System-like systems) form. 
This question is the subject of renewed interest as the number of known exoplanets continues to grow.

Surveys continue to reveal a huge diversity in exoplanet properties (see e.g., \cite{gaudi2021} for a recent review). 
New missions and facilities continue to expand the sampled parameter space, probing lower-mass planets and larger separations ($>1$~AU) from host stars. 
A model of planet formation that can reproduce the observed diversity requires considering the range of environments in which most stars and therefore most planets are born.

Making the direct connection between the observed demographics of exoplanets and their birth environments is difficult as most exoplanets are observed around mature stars. 
Few forming exoplanets have been directly detected in protoplanetary disks \cite{keppler2018,haffert2019}. 
Exoplanet systems have been detected in young open clusters, but with typical ages $>100$~Myr (e.g., \cite{fujii2019}), which is long after planet formation has completed.

While the community works to close this gap, 
plenty of progress can be made by considering the impact of the large-scale birth environment on exoplanets. 
Such outside-in models start with the observed characteristics of star-forming environments and model their impact on planet-forming disks.  
Many of the large-scale characteristics to consider for exoplanet formation are similar to the questions posed for the Solar System. 
This includes quantifying how feedback from the star-forming environment affects disk lifetimes and the architecture of the resulting planetary systems via external photoevaporation and tidal interactions (including stellar fly-bys). 
Enrichment with short-lived radioactive isotopes may play an important role in regulating the water budget of Earth-like (terrestrial) planets  \cite{lichtenberg2019}. 
Connecting these things to exoplanet demographics requires considering the diversity of star-forming regions as planet-forming environments.

Recent years have seen significant observational investment in nearby low-mass regions where high-resolution studies of many low-mass stars is possible (see e.g., reviews from \cite{manara2022,miotello2022}). 
However, observations of the Milky Way and other galaxies suggest a cluster mass function of the form $dN/dM \sim M^{-2}$.  
This means that most stars ($\gtrsim 50$\%) form in regions forming one or more high-mass stars -- star-forming complexes as massive (or more) than the Orion Nebula Cluster (ONC), itself a respectable 2000\,M$_\odot$. 
Higher-mass regions may have more -- and more intense -- ionizing feedback. 
Newly synthesized elements released in the deaths of high-mass stars, including short-lived radioactive isotopes, may pollute nearby planet-forming systems.  
More stars in high-mass regions may live in high-density environments. 
Finally, the large-scale characteristics of star-forming environments are not static. 
Understanding the evolution of star-forming regions is key to quantifying the range and impact of the local conditions a typical star experiences over its lifetime.

The density of the local star-forming environment is one of the key parameters that regulates local conditions for star- and planet-forming systems. 
The impact of external illumination from nearby high-mass stars depends on distance as $r^{-2}$. 
Tidal interactions have a steeper $r^{-3}$ dependence. 
Relatively modest changes in the typical distances between stars can therefore have a significant impact. 
Regions with high densities have shorter dynamical times and thus erase substructure and mass segregate more quickly. 
Lower density regions will retain their substructure for longer and the high-mass stars will not necessarily be centrally concentrated or clustered together at all. 

Density is not static and the density evolution of the region is key to determining which feedback mechanism dominates at a given time. 
On a global scale,  
dynamical evolution affects the amount of time that planet-forming disks are in a star-forming region. 
Determining disk lifetimes, especially in feedback-dominated regions, and comparing them to the timescale for planet formation is key to determine how important external feedback in shaping the outcome of planet formation.

Gas is an important component of the mass budget and the evolution of star-forming regions. 
In addition to impacting stellar dynamics by changing the gravitational potential (if, for example, gas is rapidly expelled from young clusters by feedback from winds and supernovae), the remnant gas in the region can protect disks from the radiation of nearby high-mass stars. 
Direct enrichment of molecular gas has also been proposed as an alternate pathway to enrich planet-forming material with short-lived radioactive isotopes (e.g., \cite{gounelle2012,gounelle2015}).  
Determining how long gas remains in region and its star-forming potential is therefore an important constraint.

In this review, we consider the large-scale environment of the star-forming complexes where most stars are born and how environmental influences affect the observed demographics of exoplanets.  
Dynamics in young star-forming complexes regulate the direct impact of photoevaporation and  gravitational interactions -- each topics that get their own chapters in this volume. 
In Section~\ref{s:definition}, we start by establish a working definition for `clustered' star formation to frame how that environment affects planet formation. 
We discuss the dynamical evolution of clusters in Section~\ref{s:dynamical} and simulations/models of the evolution of star clusters and how cluster evolution impacts planet formation in Section~\ref{s:theory}. 
In Section~\ref{s:obs} we review recent observational results on the evolution of star clusters and the impact of feedback on populations of stars/disks in clustered regions. 
We review efforts to connect observed exoplanet demographics with environmental effects in Section~\ref{s:connections} before discussing upcoming facilities and future research directions in Section~\ref{s:future}. 
Finally, we provide a brief summary in Section~\ref{s:conclusions}.

\section{`Clusters' as short-hand for `stars forming near each other'}\label{s:definition}

The claim that `most stars form in clusters' is often repeated, 
but there is not a single agreed-upon definition for what constitutes a cluster. 
Within astronomy, different communities adhere to different definitions and methods depending on whether the goal is to identify systems that are 
gravitationally bound 
or
those that are statistically clustered, as identified by algorithms like Density-Based Spatial Clustering of Applications with Noise (DBSCAN) \cite{ester1996}.

The most common reference for the claim that `most stars form in clusters' is the seminal review of \cite{lada2003}. 
From a census of star formation in the Solar neighborhood (within $\sim 2$~kpc of the Sun), they argue that 90\% of star formation happens in clusters. 
This definition includes both bound and unbound systems. 
\cite{lada2003} define a cluster as a group of stars that: 
(1) are physically related;  
(2) would be stable against tidal disruption if in virial equilibrium; and 
(3) have enough members so that evaporation time (i.e., when all members ejected by internal stellar encounters) is  $>10^8$~years (the lifetime of open clusters). 
These criteria are met for a collection of $N \geq 35$ stars or cluster with a stellar mass density $>1$~\Msun~pc$^{-3}$. 
For comparison, the average stellar mass density in the Solar neighborhood is $\sim 0.02$~\Msun~pc$^{-3}$ (assuming an average stellar mass of $\sim 0.2$~\Msun\ and using the observed number density of $\sim 0.1$~pc$^{-3}$; \cite{henry2018}).

Identifying whether a collections of stars is gravitationally bound is difficult to establish observationally (e.g., Wd2; \cite{portegies-zwart2010,zeidler2021}).  
The situation is even less clear in the star-forming stage. 
The surface density distribution of young stellar objects (YSOs) in nearby ($<500$~pc) regions show a smooth distribution \cite{bressert2010}, implying that there is no bimodal distribution to star formation, e.g. `clustered' versus `isolated', but rather that star formation results from a 
continuum of densities.

For planet formation, the key is that stars form in close proximity to other stars. 
Most stars form in regions with higher densities than the field \cite{wright2022}. 
This will affect the incident radiation field (e.g., \cite{lee2020}) 
and whether planet-forming disks are close enough to a source of short-lived radioactive elements to be enriched (e.g., \cite{looney2006,ouellette2007}). 
Whether the system remains bound beyond $\sim 10$~Myr is less important as planet formation is largely done by this point, although dynamical interactions after this time can alter the architectures of planetary systems (e.g., \cite{cai2019,winter2020}). 
Finally, while 90\% of stars form in clusters in the \cite{lada2003} definition, only $\leq 4-7$\% of these survive as bound clusters after the embedded phase. 
Clusters are either unbound during their formation (e.g., \cite{goodwin2006}) or were not bound in the first place (e.g., \cite{gouliermis2018}). 
The dissolution of previously clustered regions is a subject of great interest to connect birth conditions to demographics of exoplanets (and binary stars) observed in the field.

For this review, we consider stellar clustering in the statistical sense that stars form near other stars. 
This includes both bound and unbound systems. 
We consider how varying degrees of clustering or, more accurately, how different densities in clustered environments affect planet formation. 
%
At modest densities of a few stars\,pc$^{-3}$, aggregates are a factor $\sim$10 more dense than the field \cite{henry2018}. 
At these densities, external photoevaporation already affects disks \cite{facchini2016,nicholson2019}. 
%
At densities an order of magnitude higher, $\sim$100 stars\,pc$^{-3}$, planetary systems are disrupted by dynamical interactions \cite{adams2006,parker2012a}.
%
At higher densities of $\gtrsim 10^4$~stars\,pc$^{-3}$, 
dynamical encounters truncate disks (e.g., \cite{rosotti2014,vincke2016}).  
%
Additional considerations such as the virial state of the cluster or degree of substructure play a smaller role than density.

The cluster mass function determined from observations of the Milky Way and nearby galaxies  
is of the form  $dN/dM \sim M^{-2}$ \cite{lada2003,porras2003,chandar2009,fall2010,fall2012,mok2020}. 
The same form of the cluster mass function is found for small and large clusters although these are measured in separate surveys with samples that do not overlap 
(and include both bound and unbound clusters; see discussion in \cite{adamo2020}).  
Nevertheless, if we assume the same normalization, 
this implies that 1/2 -- 2/3 of stars form in clusters larger than the ONC (as pointed out by \cite{dukes2012}). 
These regions typically host one or more high-mass stars ($\geq 10$~\Msun). 
This proximity to high-mass stars leads to external photoevaporation of protoplanetary disks and makes possible the enrichment of planet-forming material with short-lived radioactive isotopes synthesized and ejected in the death of high-mass stars \cite{johnson2019}. 
Even a single high-mass star can have a big influence -- recall that luminosity scales with mass as $L \propto M^{3.5}$ (although note that this flattens to $L \propto M$ for masses $\gtrsim 35$~\Msun; see e.g., \cite{chandrasekhar1939}). 
Both external photoevaporation  and enrichment can have a significant impact on planet formation even in regions that are not gravitationally bound.

While high-mass star-forming complexes like the Carina Nebula are relatively rare, the sheer number of stars that they form is comparable to the combined contribution of lower-mass regions. 
A single Carina-like region is forming more stars than nearby regions like the Gould Belt clouds \emph{combined} \cite{dunham2015,povich2011}.  
Note that a cluster mass function of the form $dN/dM \sim M^{-2}$ implies that there are more low-mass clusters than high-mass, consistent with $\sim 1/3 - 1/2$ of stars forming in low-mass regions.
Existing observations overwhelmingly target these low-mass regions but these conditions are only `typical' for about half of star and planet forming systems.
For this reason, we argue that more attention should be paid to low-mass stars -- the most likely planet hosts -- forming in high-mass star forming regions.
These conditions are not sampled in relatively quiescent nearby regions (although feedback can play a role there too, see e.g., \cite{haworth2017}).  
Sampling high-mass star formation beyond the Orion star-forming complex requires pushing to the larger ($>1$~kpc) distances required to probe these `more extreme' environments. 
As we discuss in Section~\ref{s:future}, upcoming facilities will make these kinds of observations routine.

If cluster disruption and dispersal were independent of mass, then high-mass clusters would be the dominant contributors to the field population of stars -- where most known planets have been found so far. 
However, tidal stripping from the Galaxy depends on environment and will affect low-mass clusters more strongly \cite{boutloukos2003}, so the contribution of different star-forming regions to the field population may be biased to lower-mass regions.

Finally, a small fraction of stars are born in and remain in long-lived, gravitationally-bound clusters. 
The most extreme examples of high-density, gravitationally-bound clusters are globular clusters. 
In these more extreme regions, planet formation may be altered or suppressed altogether by the same mechanisms that make planet formation a challenge in less dense clusters -- disk destruction by external irradiation and dynamical encounters (e.g., \cite{adams2006,portegies-zwart2016}). 
Planets that do form are likely to be dynamically ejected (e.g., \cite{armitage2000,bonnell2001_dynamics,davies2001,spurzem2009,cai2019}). 
The more extreme conditions in globular clusters may allow 
alternate pathways for planet formation altogether, such as gravitational instability in a circumbinary disk around evolved stars \cite{beer2004_notss,beer2004}. However, only one planet has been detected in a globular cluster (in M4 -- one of the first exoplanet detections; see \cite{backer1992,sigurdsson1992,backer1993,thorsett1993,rasio1994,sigurdsson1993,sigurdsson1995,thorsett1999}) and
no transiting exoplanets were detected in the globular clusters 47~Tucanae \cite{gilliland2000,weldrake2005} or NGC~6397 \cite{nascimbeni2012}. 
It is unclear whether the apparent paucity of exoplanets in globular clusters reflects suppressed formation due to intense stellar feedback, the lower metallicity of globular clusters (planet occurance is positively correlated with stellar metallicity, \cite{fischer2005,johnson2010}), or dynamical ejection of planets in these dense systems. 

All of these factors likely contribute to the low frequency of exoplanet detections in globular clusters. 
Less clear is the fraction of planetary systems that are born in globular clusters in the first place. 
Milky Way star forming regions do not sample the low metallicities nor the high masses and densities typical of globular clusters, so there are few observational constraints on the birth conditions. 
Extrapolating from Galactic clusters may give misleading estimates if exotic pathways, such as formation in the circumbinary disk around an evolved star, contribute significantly. 
On the observational side, the
statistical evidence for a  lower planet incidence in globular clusters compared to the field is not conclusive \cite{masuda2017}. 
However, these challenging environmental conditions limit the likelihood that any planets in globular clusters are habitable planets 
\cite{kane2018} so we do not consider them further in this review.

\section{The dynamical evolution of star clusters}\label{s:dynamical}

\begin{figure}[h]%
\centering
\includegraphics[width=1.0\textwidth]{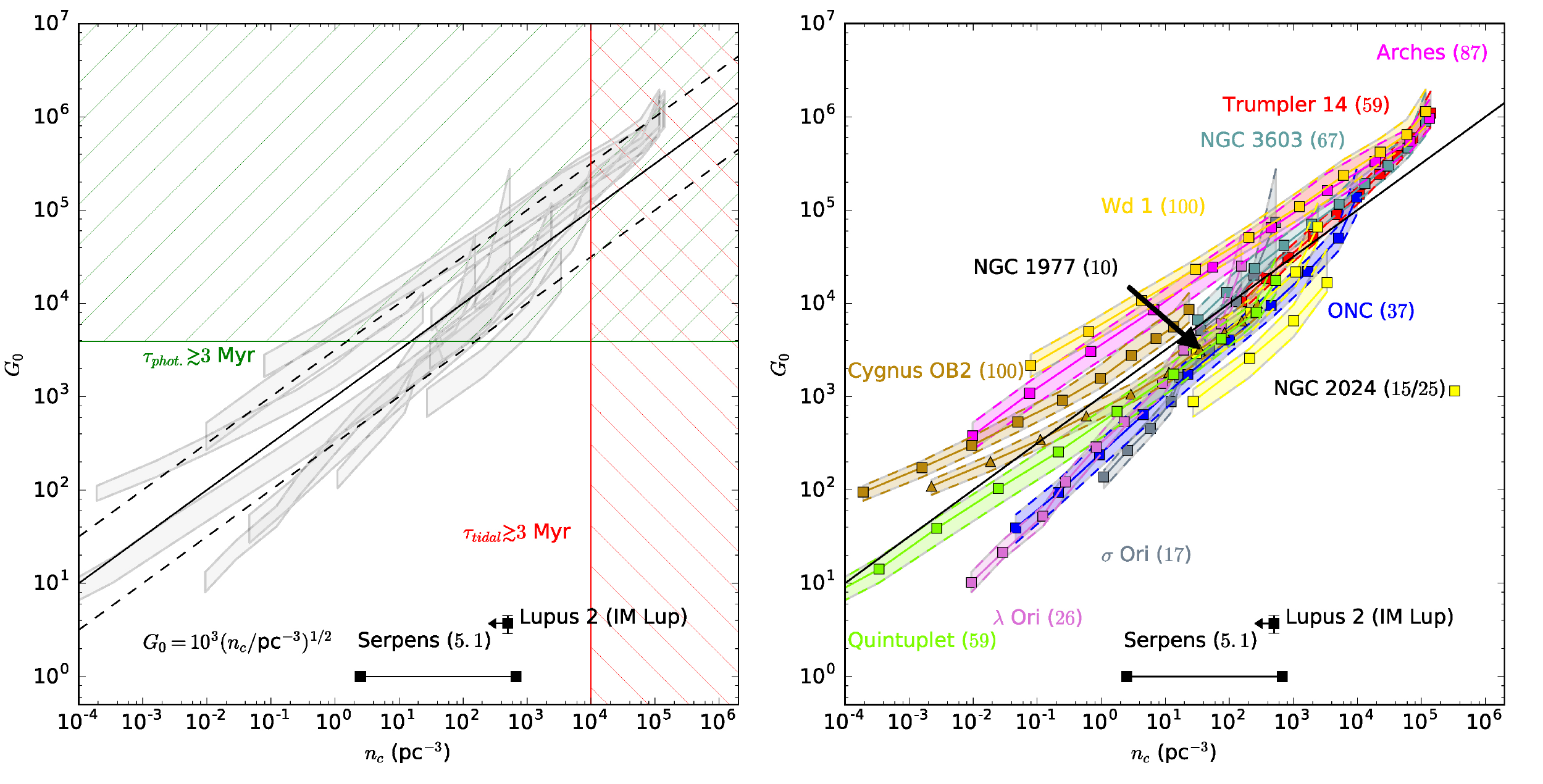}
\caption{Density is one the central determinants of the role of feedback in star/planet-formation, as illustrated by these figures from \cite{winter2018}. 
\textit{Left:} External photoevaporation  begins to impact cluster members at lower densities than tidal interactions.  
\textit{Right:} Radial-averaged densities of high-mass star-forming regions shown on the same scale for comparison. 
\textit{\textcopyright 2018 The Author(s) Published by Oxford University Press on behalf of the Royal Astronomical Society. This article is published and distributed under the terms of the Oxford University Press, Standard Journals Publication Model (https://academic.oup.com/journals/pages/about\_us/legal/notices)}.
}\label{fig:density}
\end{figure}

As discussed in the previous Section, density is a key parameter that determines how the clustered environment affects planet-forming systems. 
An example of the density dependence of different feedback mechanisms is provided in \cite{winter2018} (see Figure~\ref{fig:density}). 
\cite{winter2018} compared real cluster environments  
with the expected impact of external photoevaporation and tidal truncation.   
Radial averages in several famous star-forming clusters sample roughly 10 orders of magnitude in cluster density. 
Results from this work (and earlier studies, e.g. \cite{scally_clarke2001,adams2010}) established the density at which tidal interactions become important ($\gtrsim 10^4$~pc$^{-3}$) and demonstrated that external photoevaporation affects disks at much lower densities, $\sim$few~pc$^{-3}$. 
As a result, external photoevaporation impacts a much larger fraction of stars. 
\cite{winter2018} conclude that, where present, external photoevaporation is a dominant effect, even in regions with sufficiently high densities for tidal truncation to play a role.

The density of star clusters is not constant. 
How density evolves regulates when various feedback mechanisms have their strongest impact on planet-forming systems. 
\cite{nicholson2019} showed that photoevaporation has a stronger impact on regions that are initially low density but are undergoing collapse to higher densities. 
In contrast, \cite{vincke2016}
argue that cluster dynamics are most important for determining disk sizes through dynamical interactions. 
Density evolution in the first $\sim 10$~Myr will have the strongest impact on planet formation. 
Subsequent density evolution may enhance the rate of dynamical encounters that will sculpt the architecture of exoplanet systems. 
Understanding density evolution is therefore key to constrain which forms of feedback are important at which time[s].

Based on the high observed fraction of stars born in clusters reported by \cite{lada2003} but the low fraction of bound clusters that remain after $<$10~Myr, 
a mechanism is needed to unbind clusters if they are born bound. 
\cite{kroupa2001,goodwin2006} proposed that rapid expulsion of gas would lead to a sudden density drop that would unbind the cluster, leading to expansion and dissolution. 
In this case, loose OB associations are the expanded remnants for formerly bound clusters. 
There are a few challenges to this picture, including the timescales involved. 
Feedback mechanisms like supernovae which provide the impulse to rapidly remove remaining cluster gas happen late ($\gtrsim 10$~Myr) compared to the observed dispersal time of clusters ($<$10~Myr). 
Observations of the gas reservoir, particularly in high-mass star-forming regions, increasingly indicate that gas exhaustion not gas expulsion ends star formation 
(e.g., \cite{ginsburg2016,watkins2019}).

More recent work on the structure and kinematics of OB associations demonstrates that they are not the expanded remnants of previously high-density clusters. 
Two structural parameters indicate the degree of dynamical processing: mass segregation and the degree of substructure. 
\cite{scally_clarke2002,goodwin_whitworth2004} showed that substructure only ever decreases. 
Highly substructured regions are therefore dynamically young. 
\cite{parker2014} find that lack of mass segregation characteristic of unbound regions, whereas mass segregation tends to increase in bound clusters. 
\cite{wright2014} demonstrated significant spatial substructure and minimal mass segregation in Cyg~OB2 -- both indications that the region is dynamically young. 
This was confirmed with subsequent kinematic analysis. 
\cite{wright2016} find no evidence for global expansion in Cyg~OB2 but significant kinematic substructure in the region that appears to be close to virial equilibrium. 
Both of these facts are in contradiction to the expectations of gas expulsion.

Since the work of \cite{wright2014,wright2016}, \emph{Gaia} has provided proper motions for kinematic analysis of a large number of star-forming regions. 
Studies of young star-forming clusters \cite{kuhn2019} and young open clusters \cite{bravi2018} generally find that clusters are expanding. 
\cite{bravi2018} find that the observed expansion is slower than expected from simulations of gas expulsion \cite{goodwin2006}. \cite{parker_wright2016} show that the type of expansion observed by \cite{bravi2018} could be a signature of cool collapse, with the virial state of the region fluctuating around equilibrium, but ``freezing in" a supervirial velocity dispersion. 
\cite{ward2020} note that the lack of correlation between radial velocity and radius argues against expansion from a previous high-density state.

The interpretation of kinematic data remains debated, and discrepancies between radial velocity and proper motion data are observed (see discussion in e.g., \cite{bravi2018}). 
Both expansion and collapse have been claimed in some regions and, in some cases, from the same data. 
\cite{kuhn2019} found evidence for mild expansion of the ONC using \emph{Gaia} DR2 and no evidence for hierarchical assembly in the 28 clusters they consider. 
In contrast, 
\cite{tobin2009} interpret their radial velocity data of the ONC to be consistent with models of cold collapse. 
Similarly, \cite{wright2019} find that the mass-dependent proper motion velocity dispersion seen in NGC~6530 is consistent with expectations for a cluster assembling hierarchically, in contrast to \cite{kuhn2019} who find a radially-dependent expansion velocity. 
Breaking degeneracies between these interpretations requires combining 
analysis of spatial structure,  
6D kinematics (x, y, z, v$_x$, v$_y$, v$_z$), and 
models (see \cite{parker_wright2016}). 

Furthermore, \emph{Gaia} data have been used to constrain the initial densities of star-forming regions by determining the number, and velocity distribution, of runaway stars (stars moving faster than 30\,km\,s$^{-1}$) and slower, walkaway stars (stars moving $>$5\,km\,s$^{-1}$). Before \emph{Gaia}, walkaway stars were impossible to trace back to star-forming regions, but many more walkaway stars are produced than runaways as their production rate depends on stellar density \cite{schoettler2019}. These new results from \emph{Gaia} corroborate the earlier constraints on the initial conditions of star-forming regions from spatial structure and mass segregation \cite{parker2022}.

Results from \emph{Gaia} have been transformative in many ways, but because it is an optical mission, it is limited to mostly unobscured (and therefore older) regions. 
However, as \cite{nicholson2019} and others highlight, the \emph{initial} density is a key parameter regulating how the larger environment affects planet-forming disks. 
Probing younger regions and identifying initial conditions is an important direction for the future. 
Upcoming IR-optimized facilities (see Section~\ref{s:future}) will help. 
The sensitivity and angular resolution of existing facilities like the Atacama Large Millimeter/Sub-millimeter Array (ALMA) already permit studies of embedded low-mass stars in high-mass regions \textbf{(e.g., \cite{yusef-zadeh2017,walker2021,redaelli2022})}.  
Much less is known about the initial distribution of low-mass sources in high-mass regions compared to the well-studied local clouds, but this is key to constrain the impact of feedback in the early stages.

Focusing on earlier stages also requires considering the coevolution of stars and gas. 
As an example, consider the Orion star-forming complex. 
Gas kinematics suggest large-scale filament collapse is feeding Orion A \cite{fallon1977,hacar2017}. 
This is broadly consistent with gravity-only 
simulations of cold collapse from \cite{kuznetsova2015,kuznetsova2018} and observations of the radial velocities of stars in the ONC \cite{furesz2008,tobin2009}. 
Within the ONC itself, there is evidence for multiple populations \cite{beccari2017,jerabkova2019} with the 
youngest stars concentrated in the cluster center.  
\cite{winter2019} proposed that this may explain the ongoing presence of disks in the ONC despite their high observed photoevaporation rates (for which disks in a static, single-age cluster would already be completely destroyed; see Section~\ref{s:obs} for further discussion).

Core-halo age gradients with younger stars concentrated in the center of clusters have also been suggested in the ONC and NGC~2024 using an X-ray chronometer \cite{getman2014_core_halo}. 
Using this method, age gradients in several star-forming regions have been suggested \cite{getman2014,getman2018}. 
Such core-halo age gradients are a prediction of  hierarchical collapse models \cite{vazquez-semadeni2017,vazquez-semadeni2019}.

We use Orion to illustrate how the large-scale evolution can affect disks in star-forming regions -- in this case, delivering young (disk-bearing) stars into the intense radiative environment near the high-mass stars. 
However, we caution that the interpretation of the kinematics of stars in the ONC and the Orion~A cloud are the subject of on-going debate \cite{muench2008} with arguments for 
neither expansion or contraction \cite{dzib2017} and multiple studies finding evidence for 
expansion \cite{dario2017,kounkel2018,kuhn2019}.   
The large-scale complex may be affected by larger scale influences as suggested by the combined motion of multiple clusters associated with the region \cite{grossschedl2021,kounkel2021}. 
Nevertheless, Orion provides a useful illustration of how large-scale multi-wavelength observations with excellent spatial and velocity resolution can be used to unpack star-forming regions as the 
dynamic birthplaces of planets.

\section{Models}\label{s:theory}

\subsection{Cluster evolution}\label{ss:models_evolution}

Hydrodynamical simulations provide insights into the early stages of star formation, but cannot follow the long-term evolution. Typically, simulations can either resolve brown dwarfs and binaries (e.g., \cite{bate2009,bate2012}) 
or 
high-mass stars and the impact of their feedback (e.g., \cite{dale2011,dale2012,dale2013,dale2013_winds}), but not both.  
The newest generation of simulations now include multiple forms of stellar feedback -- jets, winds, radiation, supernovae (e.g., STARFORGE; \cite{grudic2021}). 
These simulations model star formation in giant molecular clouds, sampling the high-mass star-forming regions in which most stars form. 
At the same time, they resolve individual stars, 
allowing the direct connection between feedback in the larger star-forming environment to its impact on individual star- and planet-forming systems. 
Such simulations promise to be important for understanding the dynamical evolution of clusters during gas removal due to pre-supernova feedback as well as providing 
initial conditions to use for higher-resolution simulations of individual planet-forming disks.

There have been a few efforts to combine long-term N-body simulations with a gas potential (e.g., \cite{hubber2013,sills2018}) and with hydrodynamically evolving disks (e.g., \cite{rosotti2014}). 
Such efforts require adopting realistic spatial and kinematic initial conditions, but this is a challenge given 
the mismatch in the angular resolution of available observations of stars and gas. 
Wide-field surveys like the 
Massive Young Star-forming Complex Study in Infrared and X-ray (MYStIX; \cite{feigelson2013}) 
and 
Star Formation In Nearby Clouds (SFiNCs; \cite{getman2017}) provide distributions of low-mass stars on scales of $<1^{\prime\prime}$. 
In contrast, wide-field surveys of molecular gas distribution in star-forming regions is typically done with single dish telescopes with angular resolution $\sim 30^{\prime\prime}$, $1-2$ orders of magnitude more coarse than the stellar measurements. 
Realistic discretization for grid-based simulations is a significant challenge. 
Observational improvements are on the horizon with the development with large aperture telescopes like the Atacama Large Aperture Submillimeter Telescope (AtLAST; \cite{klaassen2019_atlast}) which will provide an order of magnitude improvement in survey resolution (see Section~\ref{s:future}).

\subsection{Cluster impact on disks}\label{ss:models_disks}

Most studies of planet formation in the cluster environment focus on the direct impact of external feedback on protoplanetary disks. 
With growing evidence that planet formation happens early (e.g., \cite{segura-cox2020}), there is growing urgency to consider earlier stages. 
If planet formation is well advanced before the disk is exposed to feedback from the nearby high-mass stars, then feedback will play a much smaller role in shaping the outcome of planet formation. 
One of the key questions now is how quickly planets form and how soon feedback affects ongoing planet formation to determine how long, if ever, these two processes are in direct competition.

When considering these earlier stages, the co-evolution of gas and dust is very important beyond its role in regulating the dynamical evolution of the cluster. 
Remaining cloud material shields disks from the full strength of the radiation field of nearby high-mass stars, reducing the incident radiation and prolonging the lifetime of the disk. 
\cite{van_terwisga2020} proposed this as one explanation of the two population of disks observed in NGC~2024. 
Simulations of feedback in a Carina-like high-mass star-forming region from \cite{qiao2022} find that shielding by remnant gas can protect disk for $<0.5$~Myr but that this may be enough to allow for significant evolution of the solids, helping ensure the mass budget for terrestrial planet formation. 
Within 2~Myr, the majority of disks in the region are affected by external photoevaporation, demonstrating that either planet formation must happen quickly or that giant planet formation may be suppressed in high-mass star-forming regions.

Several studies have specifically modelled the survival and enrichment of planet-forming disks in dynamically evolving clusters. 
\cite{lichtenberg2016}
modelled direct enrichment of planet-forming disks with short-lived radioactive isotopes from supernovae in clusters of different morphology. 
They find a wide variation in the level and likelihood of enrichment, suggesting that Solar-System-like levels of enrichment may be environment dependent.

In light of this tension between disk lifetime and the time for high-mass stars to evolve off the main sequence and explode as supernovae,  
\cite{nicholson2017} explored 
supernova enrichment in low-mass clusters. 
They find that direct enrichment in clusters with one or two $>20$~\Msun\ stars is as likely as in higher-mass clusters 
because disks are destroyed faster in higher-mass clusters.  
Indeed, rapid disk dispersal is a challenge.  
\cite{nicholson2019}
showed that disks are rapidly dispersed on short timescales, with half dispersed in 2~Myr in regions with densities as low as $\sim 10$~\Msun~pc$^{-3}$. 
Through a series of N-body simulations sampling a range of initial densities 
as well as cluster structures and bulk motions, they demonstrate that the \emph{initial} density is the most important parameter regulating the impact of external photoevaporation on protoplanetary disks. 
More disks are photoevaporated in regions undergoing collapse than those in virial equilibrium or still expanding. 
In lower density regions ($\sim 10$~\Msun~pc$^{-3}$), substructure increases the rate that disks are destroyed, although other effects dominate at higher densities ($\gtrsim 100$~\Msun~pc$^{-3}$). 
They conclude that either: 
planet formation must happen quickly;  
giant planet formation will be confined to smaller radii than the orbit of Neptune; or 
that disk evaporation rates are overestimated (in contradiction to recent models; e.g., \cite{haworth2018_fried}). 

Photoevaporation models strongly depend on assumptions about the initial disk radii as it primarily impacts the outer disk. 
Material at smaller disk radii and around higher mass stars is more tightly graviationally bound and therefore less vulnerable to removal by external irradiation \cite{haworth2018_fried}. 
For example, we consider 
a 1\,M$_\odot$ star surrounded by a disk with a radius $r_{\rm disk} = 100$\,AU and a mass $M_{\rm disk} = 0.1$~M$_\odot$ in a $G_0 = 10000$ radiation field (common in clusters containing high-mass stars). 
The \texttt{FRIED} grid of models of externally evaporating protoplanetary disks \cite{haworth2018_fried} predicts a photoevaporation rate of $3 \times 10^{-6}$~M$_\odot$\,yr$^{-1}$, so our example disk will be destroyed within a Myr. 
If the initial disk radius were instead 30\,AU, the \texttt{FRIED} grid predicts a much lower photoevaporation rate of $4 \times 10^{-8}$~M$_\odot$\,yr$^{-1}$, corresponding to a survival time of several Myr. 
In practice, the resulting disk lifetime depends on the viscosity of the disk (which may help explain why the hot inner disks also disappear more quickly in feedback-dominated regions; see Section~\ref{s:obs}). 
Nevertheless, this example illustrates that 
our understanding of the distribution of initial disk radii is crucial; if the initial radii are several 100s AU (as in the DSHARP survey; \cite{andrews2018}), we would expect photoevaporation to dominate over other disk destruction mechanisms.

\cite{concha-ramirez2019}
also examined how external photoevaporation in the clustered environment constrains the timescale for planet formation 
in simulations that include dynamical interactions and external photoevaporation. 
They find that $\sim 50$\% of disks are destroyed in $<2$~Myr in clusters with mass densities $\rho \sim 50$~\Msun~pc$^{-3}$. 
At higher mass densities, $\rho \sim 100$~\Msun~pc$^{-3}$, the fraction of disks destroyed within 2~Myr rises to 80\%. 
Follow-up work in \cite{concha-ramirez2021}
demonstrated that disk destruction depends on the density of the region, in agreement with other studies. 
They find an order-of-magnitude variation in disk masses with order-of-magnitude variation in region density. 
They demonstrate that disk masses determined from the mm continuum flux in nearby high-mass star-forming regions are broadly consistent with their models.

Studies of planet-forming disks in the context of star-forming clusters are broadly in agreement. 
Disks are rapidly dispersed by external photoevaporation. 
More disks destroyed earlier in regions with higher densities and stronger UV fields. 
The fraction of disks destroyed varies between studies, but most agree on $>50$\% in $\lesssim 2$~Myr. 
Together, this strongly suggests that planet formation must happen early ($<1$~Myr).

A growing number of studies are trying to connect the influence of clustered star-forming environment with the architectures of the resulting planet populations (e.g., \cite{ndugu2018,flammini_dotti2019,flammini_dotti2020,wang2020_clusters,stock2020,wang2020_hotJupiters,ndugu2022,stock2022}). 
Many of these focus on dynamical interactions that sculpt already formed planetary systems. 
However, a few look at the impact of the cluster environment during planet formation. 
\cite{ndugu2018} synthesize planet populations from a range of initial conditions for the planetary system and the clustered environment. 
They show that background heating in the cluster environment increases the temperature in outer parts of the disk, suppressing the formation of cold Jupiters to levels consistent with observations.  
Disk truncation due to stellar encounters may also reduce the fraction of cold Jupiters formed in clustered regions \cite{ndugu2022}. 
At the moment, these simulations treat different forms of feedback in isolation. 
As \cite{ndugu2022} urge, future work should consider the contributions of multiple feedback mechanisms acting in concert -- external photoevaporation in addition to background heating and tidal interactions.

\section{Observational signatures}\label{s:obs}

\begin{figure}[h]%
\centering
\includegraphics[width=0.9\textwidth]{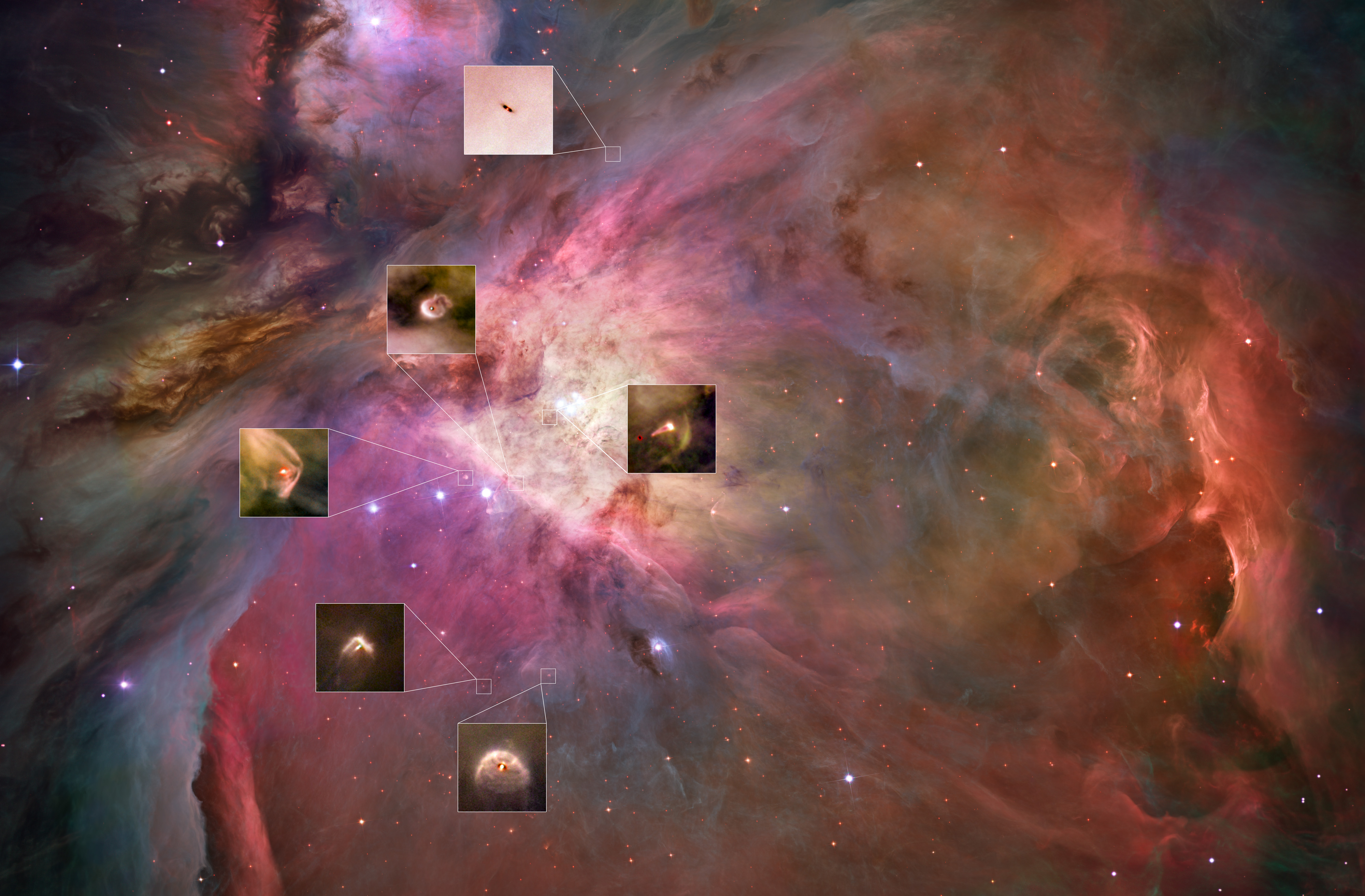}
\caption{
An \emph{HST} image of the ONC with postage-stamp images highlighting the famous proplyds. 
\textit{Credit: NASA, ESA, M. Robberto (Space Telescope Science Institute/ESA), the Hubble Space Telescope Orion Treasury Project Team and L. Ricci (ESO)}.
}\label{fig:proplyds}
\end{figure}

Some of the first and most spectacular evidence of the impact of feedback from high-mass stars on planet-forming disks were images of the photoevaporating protoplanetary disks (proplyds) in Orion from the \emph{Hubble Space Telescope (HST}; see Figure~\ref{fig:proplyds}) \cite{odell1993,odell1994}. 
External UV radiation from nearby high-mass stars illuminates disks, driving a neutral photoevaporative flow that is then ionized. 
This creates a cometary or tear-drop shape with a rounded head pointed toward the dominant ionizing source and a tail that extends in the opposite direction. 
For nearby regions, the characteristic tear-drop shape 
can be identified in diffraction-limited images. 
Confirmed proplyds are detected near high-mass O- and B-type stars that drive high photoevaporation rates in the disks, 
$10^{-8} - 10^{-6}$~\Msun~yr$^{-1}$ \cite{johnstone1998,storzer1999,henney1998,henney1999,henney2002}. 
With these high evaporation rates, proplyds should be rapidly dissipated.

There have been multiple attempts to visually identify proplyds in star-forming regions other than the ONC. 
Most studies use narrowband images of  hydrogen recombination lines (i.e.\ H$\alpha$; Paschen-$\alpha$) to search for the characteristic tear-drop morphology. 
With this approach, candidate proplyds have been identified in 
NGC~1977 \cite{kim2016}; 
NGC~2024 \cite{haworth2021_proplyds}; 
Cyg~OB2 \cite{wright2012};  
Carina \cite{smith2003}; and  
NGC~3603 \cite{brandner2000}. 
Some candidate proplyds are $20-30\times$ the size of the ONC proplyds \cite{brandner2000}, calling into question their classification. 
Mass estimates for a few objects derived from millimeter measurements of the gas and dust are on the order of stellar masses, not disk masses (e.g., \cite{sahai2012,reiter2020_tadpole}). 
Despite their similar morphology, many candidates may instead be evaporating gaseous globules (EGGs).

A few challenges frustrate efforts to identify proplyds in more distant high-mass regions. 
Many proplyd candidates are located several pc from the main ionizing sources, unlike the ONC proplyds which are detected within $\sim 0.1$~pc of $\theta^1$C. 
True proplyds, analogous to those in the ONC, are unlikely to be spatially resolved in more distant ($>2$~kpc) regions. 
In their seminal paper, \cite{johnstone1998} show how the stand-off distance, $r_{\mathrm{IF}}$, the distance between the disk and the ionization front of the neutral wind flowing from it, depends on the incident radiation. 
They show that $r_{\mathrm{IF}}$ increases with distance from the ionizing source[s] but decreases as the radiation field gets stronger. 
Specifically, 
$r_{\mathrm{IF}} \propto \Phi^{-1} d^2$
where 
$r_{\mathrm{IF}}$ is the stand-off distance; 
$\Phi$ is the incident ionizing flux; and 
$d$ is the distance from the ionizing source.  
Typical $r_{\mathrm{IF}}$ sizes are $\sim 20-200$~AU 
\cite{johnstone1998,haworth2021_proplyds}. 
At the distance of the ONC ($\sim 400$~pc; \cite{grossschedl2018}), 
this corresponds to an angular size of $\sim 0.05^{\prime\prime} - 0.5^{\prime\prime}$. 
From more distant regions like Cyg~OB2 (1.33~kpc; \cite{kiminki2015}), a comparable proplyd would have an angular size of $\sim 0.015^{\prime\prime} - 0.15^{\prime\prime}$ which would be difficult but possible to resolve with current facilities. 
Expected proplyd sizes are even smaller at the distance of Carina (2.3~kpc; \cite{goeppl2022}), with ONC-like objects having sizes of $\sim 0.009^{\prime\prime} - 0.09^{\prime\prime}$. 
To make matters worse, these small sizes are optimistic upper limits given that the ionizing flux, $\Phi$, is at least an order of magnitude higher in regions like Carina \cite{smith2006_energy}, with a corresponding decrease in $r_{\mathrm{IF}}$ (photoevaporation rate of globules in Carina are estimated to be $\sim 10^{-5} - 10^{-6}$~\Msun~yr$^{-1}$ \cite{smith2004_finger,reiter2019_tadpole}). 
Spatially resolved observations are therefore likely beyond the reach of the \emph{James Webb Space Telescope} (\emph{JWST}). 
The shortest wavelength narrowband filter is Paschen-$\alpha$ at 1.87~$\mu$m, providing $0.07^{\prime\prime}$ imaging.

While the stronger radiation field reduces the likelihood that externally photoevaporating sources will be spatially resolved, stronger fields may affect a larger fraction of sources in a given region. 
As a result, interest is growing in identifying signatures of external photoevaporation that do not require spatially resolving the source morphology. 
Directions for future work include identifying photometric and spectroscopic signposts and predicting statistical prevalence in populations of different ages.

Millimeter continuum emission traces cold dust in outer disks. 
Recent measurements also suggest that disk masses are lower near high-mass stars. 
The ONC \cite{mann2010,mann2014,eisner2018} and 
$\sigma$~Ori \cite{ansdell2017}
both show structure in the disk mass distribution. 
The lowest mass disks are where the impact of external photoevaporation is strongest, within $\sim 0.5$~pc the main ionizing source. 
These radially dependent masses are not universally observed, however. 
\cite{mann2015} do not observe a trend in disk masses with distance from the O8V star in NGC~2024, in contrast to the ONC and $\sigma$~Ori. 
This was confirmed with recent ALMA observations from \cite{van_terwisga2020} who suggest that remnant cloud gas may shield disks from the harshest radiation 
in this young (0.5~Myr) region, modulating disk destruction. 
Little gas remains to shield the disks in the older ($\sim 1$~Myr) population in NGC~2024. 
As in the young regions, disks in the older region do not show a structured distribution of disk masses. 
Fewer disks are detected in the older population and their masses are lower, perhaps reflecting the more direct irradiation from the B-type star that illuminates the region. 
In young regions, remnant gas may play a more important than the earlier spectral type of the star for modulating the disk population.

Structured distributions of disks have also been suggested in higher-mass regions. 
Without high-resolution millimeter observations to measure disk masses, these studies use the presence or absence of a disk as traced by near-IR excess emission. 
Excess near-IR emission compared to the photosphere indicates the presence of a hot ($\sim 2000$~K) inner disk. 
Radial averages suggest a higher fraction of disks further from the high-mass stars (e.g., \cite{guarcello2007,guarcello2016}).  
Such clear signatures are not seen in every region, especially where the high-mass stars are more widely distributed (e.g., \cite{reiter2019}).

There are a few reasons that a structured distribution of disks may not be the best way to infer the impact of feedback on disks. 
Perhaps most fundamental is that the distribution of disk-bearing sources is seen in projection. 
Projected distances between low- and high-mass stars may substantially underestimate the true distance. 
Parallax measurements from \emph{Gaia} are clarifying the 3D distribution of stars in clusters. 
Unfortunately, precision is lowest for the faintest (lowest mass) sources but these are the most affected by external feedback.
Regions with multiple high-mass stars may not be mass segregated, so disk-destroying radiation may affect the disk from multiple directions. 
In substructured regions, there may be less disk photoevaporation (compared to a smooth region with similar density) because on average, more stars are further from the high-mass stars \cite{parker2021}. 
However, stars move. 
For a velocity of 1~\kms, a star will move 1~pc in 1~Myr 
so a star near a high-mass star now was not necessarily there in the past and will not necessarily be there in the future. 
Tracing stellar motions back in time is increasingly possible using data from \emph{Gaia} but this cannot capture the full dynamical history (e.g., \cite{arnold2017}).

External irradiation also heats disks, leading to higher temperatures that may alter the chemistry and evolution of planet-forming material.  
At the most basic level, accurate disk temperatures are crucial to correctly determine the disk mass \cite{haworth2021_temp}. 
A higher disk temperature translates to lower estimated mass under standard assumptions (e.g., Table~1 in \cite{mesa-delgado2016}). 
Using representative temperatures will be key to enable comparisons of  
disk masses and demographics within and between regions.  
External heating may also affect ice lines \cite{haworth2021_temp} with high density systems potentially heating disks to the point where the ice line disappears, suppressing giant planet formation \cite{thompson2013}.

Dynamical interactions also perturb disks with observable consequences. 
Models of the impact of stellar flybys \cite{cuello2019,cuello2020}
predict warps, spirals, truncation, and increases in the accretion rate. 
At the moment, the best observations of the impact of tidal interactions on protoplanetary disks come from nearby binary systems, for example 
the multiple flybys sculpting RW~Aur \cite{rodriguez2018} and  
tidally truncated disks in Taurus \cite{manara2019}. 
Resolving comparable features in more distant regions will be a challenge. 
The ONC is the nearest high-density region at a distance of $\sim 400$~pc \cite{grossschedl2018}. 
This is $\sim 2-3\times$ the distance of disks where detailed substructures have been resolved (e.g., \cite{andrews2018}).

\begin{figure}[h]%
\centering
\includegraphics[width=0.7\textwidth]{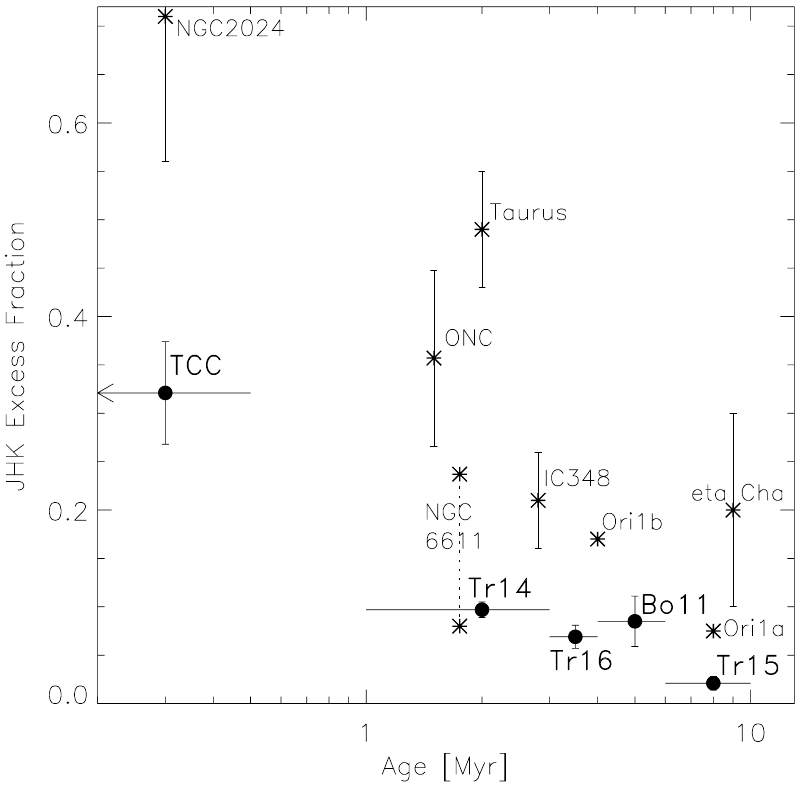}
\caption{External photoevaporation may reduce the disk lifetime in regions with multiple high-mass stars. 
Figure from \cite{preibisch2011_hawki} showing the near-IR excess fraction, indicating the presence of a warm inner disk, as a function of the region age comparing the high-mass clusters in Carina (black dots) with lower-mass clusters from \cite{haisch2001} (stars). 
\textit{Credit: Preibisch, T., et al., A\&A, 530, A34, 2011, reproduced with permission} \textcopyright\ \textit{ESO}. 
}\label{fig:disks}
\end{figure}

Going forward, we expect renewed interest in population-level studies to quantify the impact of environment. 
The seminal paper of \cite{haisch2001} showed a steady decline in the near-IR excess fraction (disks) as a function of age in nearby star-forming regions. 
\cite{stolte2010} suggested that disk lifetimes might be shorter in regions with many high-mass stars, although in these distant regions that signal was difficult to disentangle from the faster evolution of high-mass sources. 
More recent work targeting truly low-mass stars in the Carina star-forming complex support shorter disk lifetimes in higher-mass regions. 
\cite{preibisch2011_hawki} measured 
significantly lower disk fractions in the main clusters of Carina -- Tr14, Tr15, and Tr16 -- than in similarly aged regions without high-mass stars (see Figure~\ref{fig:disks}).

This result is not universal, with other studies finding no evidence that there are fewer disks near high-mass stars \cite{richert2015}. 
Larger, more uniform surveys of disks in high-mass regions will help clarify the impact of feedback on disks \textbf{(e.g., \cite{mendigutia2022})}. 
Better age estimates will also be required to accurately assess changes in disk populations and demographics with time.

\section{The role of cluster dynamics in enrichment}\label{s:enrichment}

The origin and injection of short-lived radioactive isotopes has been a subject of long-standing interest for the Solar System \cite{cameron1977,chevalier2000,gaidos2009,adams2014,fatuzzo2015,fatuzzo2022}. 
The most attention has been paid to  
\alr\ and \fer\ due to their high abundances and short radioactive decay times (half-life of 0.717~Myr and 2.62~Myr for \alr\ and \fer, respectively). 
As such, they represent a significant heat source for early geochemical evolution.  
Given the important role these elements play in the Solar System, there is ongoing interest in how frequently Solar-System analogs are enriched (e.g., \cite{gounelle2015,portegies-zwart2019}).

Determining the source of these elements provides one of the strongest constraints on the birthplace of a planetary system. 
Both elements are produced in the late stages of the evolution of high-mass stars. 
Supernova explosions are one of the largest sources of both elements. 
Wolf-Rayet (W--R) stars may also release significant amounts of \alr\ in their intense winds. 
Other sources include evolved intermediate-mass asymptotic giant branch (AGB) stars, although these are unlikely to provide enrichment to other stars in their birth cluster while planet formation is ongoing. 
Alternatively, \alr\ (but not \fer) may be produced locally within the disk by cosmic ray spallation. 
See \cite{lugaro2018} for a more complete review.

For outside-in models that seek to understand how the larger star-forming environment affects the outcome of planet formation, the key questions are 
when, 
where, and 
how long these elements are present and, 
crucially, 
whether they mix with planet-forming material.  
Fortunately, \alr\ and \fer\ emit $\gamma$-ray photons when they radioactively decay, so it is in principle possible to measure their abundance and distribution. 
Maps of the Milky Way from the \emph{COMPTEL} and \emph{INTEGRAL} $\gamma$-ray observatories (e.g., \cite{diehl1995,diehl2006_integral,bouchet2015}) reveal \alr\ to be confined close to Galactic plane and bright in high-mass star-forming regions \cite{knoedlseder1999,diehl2006}. 
The situation is more complex for \fer. 
It has a lower abundance, with an observed 
\fer/\alr\ $\gamma$-ray flux ratio of 18.4\% $\pm$ 4.2\% \cite{wang2020}. 
The Galactic \fer\ signal is too weak to constrain the spatial distribution of the emission. 
However, a comparison with the distribution of \alr\ and plausible morphologies suggests that the elements have a different Galactic spatial distribution \cite{wang2020}. 
More sensitive maps of both \alr\ and \fer\ are goals of the Compton Spectrometer and Imager (COSI; \cite{tomsick2022,beechert2022}) and will hopefully improve the global view.

The $\gamma$-ray observations indicate that the majority of Galactic \alr\ is produced by high-mass stars.  
The distribution within high-mass star-forming regions is poorly constrained as the resolution of the $\gamma$-ray observations are $\gtrsim 1^{\circ}$. 
This is unlikely to improve in the immediate future given the enormous technical challenges involved in $\gamma$-ray observations. 
Models therefore remain a key element to connect the observed large-scale distribution of \alr\ and the likelihood it enriches planet-forming material. 
One under-explored avenue is to examine the abundance of \alr\ as a function of time. 
Population synthesis modelling of well-studied regions like Orion and Carina suggest a high abundance of \alr\ that is maintained for several Myr \cite{voss2010,voss2012}. 
For regions with multiple high-mass stars, regular replenishment may sustain high levels for much longer than the 0.717~Myr half-life. 
\cite{reiter2020_al} point out that this could enable significantly more systems to be enriched. 
Models including the production and mixing of \alr\ on the cloud (e.g., \cite{fatuzzo2022}) and Galactic (e.g., \cite{fujimoto2018,fujimoto2020}) scales will provide more realistic constraints on the number of systems that are likely to be enriched.

Recent surveys of high-mass stars also provide improved inputs for modelling efforts (e.g., \cite{cjevans2011,sana2012,holgado2018,holgado2022}). 
Two key physical properties merit renewed attention. 
First, stellar rotation rates affect how much \alr\ mixes into the winds from high-mass stars (e.g., \cite{voss2009}). 
Fast rotators introduce \alr\ earlier and for lower initial masses than non-rotators. 
However, the observed rotation rates of high-mass stars are slow ($\sim 100$~\kms), more consistent with non-rotating models \cite{cjevans2020}. 
In the context of the star- and planet-forming environment, this will affect both how early and how long \alr\ is available in the region. 
A second, related consideration is that most high-mass stars are in multiple systems and the majority ($\geq 70$\%) will eventually interact \cite{sana2012}. 
Models for binary evolution are in active development.  
For \alr, recent work from \cite{brinkman2019} suggests that binary interactions will increase the contribution from $\sim 10-15$~\Msun\ stars but that higher-mass stars ($\gtrsim 30$~\Msun) will still dominate the production of \alr. 
Similar work for a broader range of elements may help further constrain which environments / stellar populations are likely to lead to enrichment with \alr\ and other crucial elements.

Additional constraints come from the observed distribution of probable \alr\ sources. 
A recent study used \emph{Gaia} data to determine the environment of Galactic W--R stars. 
Outside of the Galactic Center, only $18-36$\% of Galactic W--R stars appear to be associated with clusters, associations, and/or star-forming regions \cite{rate2020}. 
The percentage of apparently isolated W--R stars is much larger than main-sequence O-type stars. 
\cite{rate2020} estimate that 
$\sim 20$\% of the isolated W-R stars could be ejected from a clustered birth region by dynamical ejection or binary disruption. 
These results imply that a large fraction ($\sim 45$\%) of the progenitor stars form in low-density associations so that they appear isolated by the time they evolve to W--R phase. 
Any models of W--R enrichment should include the relatively small probability that a W--R star is located close to disk-bearing stars or an appreciable quantity of star-forming gas. 
For example, a study that estimated the likelihood that an AGB star pollutes a molecular cloud in the Solar neighborhood found a typical probability of $\sim 1$\% \cite{kastner1994}. 

One of the main challenges with external sources providing short-lived radioactive elements to planet-forming systems is the timescale. 
On the one hand, 
stellar evolution happens comparatively slowly with a 25~\Msun\ star taking $>7$~Myr to evolve off the main sequence and explode as a supernova. 
On the other, evidence is growing that planet-formation happens early ($<1$~Myr; see e.g., \cite{segura-cox2020}). 
\cite{dukes2012} suggest considering higher-mass clusters as more massive stars will have shorter lifetimes (and the highest yields; \cite{limongi2006,woosley2007,ekstrom2012,limongi2018,brinkman2019}), reducing the discrepancy in these timescales. 
However, increased disk destruction from the more intense radiation fields of higher mass stars may instead exacerbate the discrepancy. Enrichment of nearby star-forming gas may provide an alternate pathway, but the timescale problem remains as this still requires an evolved high-mass star located next to star-forming gas. 
Pre-supernova feedback appears to rapidly clear remnant cloud material from high-mass star-forming regions (e.g., \cite{dale2012,kruijssen2019,chevance2020}). 
Sequential star formation, as outlined by 
\cite{forbes2021}, provides another possibility.  
In this case \alr\ and \fer\ come from (primarily) supernova explosions from the Sco-Cen association trigger then enrich star-forming cores in the nearby L1688 star-forming region. 
This scenario requires multiple supernovae -- first to set star formation in motion literally and figuratively 
with the blastwave from one or more supernovae, followed by enrichment. 
Better age estimates may provide a test of this scenario as star formation directly triggered by feedback from high-mass stars is expected to lead to large ($>10$~Myr) age differences between populations \cite{parker_dale_2016}.

Other enrichment scenarios, such as cosmic ray spallation, may be less sensitive to environment. 
Recent models propose that cosmic rays accelerated in accretion shocks provide the necessary flux of energetic particles to produce \alr\ locally \cite{gaches2020}. 
In this case, more environments may be able to produce \alr-rich systems. 
An environmental dependence may persist if external cosmic rays contribute significantly to \alr\ production. 
High-mass stars themselves may be a significant source of cosmic rays \cite{aharonian2019}. 
Stellar winds and supernova explosions interacting with the surrounding medium may enhance the local cosmic ray flux  \cite{fatuzzo2006,liu2022}.  
Models including both internal and external cosmic ray sources would be interesting to explore.

As an aside, we note that an enhanced cosmic ray flux in high-mass star-forming regions would be interesting to explore more generally. 
Cosmic rays play an important role in planet-forming disks beyond element synthesis \cite{padovani2018}. 
Cosmic rays are the only source of ionization deep in molecular clouds \cite{indriolo2009} and disk midplanes, and thus enable chemistry in planet-forming disks \cite{umebayashi1981,cleeves2013,cleeves2015}.  
Changes in the ionization level due to different cosmic ray fluxes affects magnetically-mediated processes like accretion and outflow that require some ionization to couple gas to the magnetic field (e.g., \cite{offner2019}). 
Higher ionization may enhance magnetic braking which \cite{kuffmeier2020} propose as an alternate explanation of the smaller disk sizes observed in high-mass star-forming regions.

\section{Connecting birthplaces to planet properties}\label{s:connections}

As discussed in the previous section, short-lived radioactive isotopes provide some archaeological evidence for the birthplace of the Solar System and clues as to how the birth environment might affect exoplanet demographics. 
With the enhanced view of Galactic dynamics provided by \emph{Gaia}, some groups have also tried to back out the birth environment of the Solar System dynamically. 
This includes work to reconstruct dissolved clusters to identify birth location and cluster origin of the Solar System \cite{moyano_loyola2015}. 
The Galactic open cluster M67 has been proposed as an analog of the Solar System birthplace because of its similar age and metallicity. 
The cluster is sufficiently massive to have produced high-mass stars to provide enrichment with short-lived radioactive isotopes. 
\cite{jorgensen2020} determine that stars escaping from M67 can reach Solar-like orbits.  
Most clusters do not survive for Gyr, so an origin in a surviving cluster like M67 would point to an unusual birthplace for the Solar System compared to most stars. 
External photoevaporation from high-mass stars in the cluster (that also provided short-lived radioactive isotopes) may have prevented the formation of Jupiter and Saturn.

Direct observations of planets in their birth clusters remain challenging with only a few confirmed detections of accreting protoplanets (i.e., PDS~70; \cite{keppler2018,haffert2019}). 
Planets have been detected orbiting stars in young ($>100$~Myr) open clusters providing important age constraints for models \cite{quinn2012,fujii2019}. 
Planets in longer-lived $\sim$~Gyr systems like NGC~6811 \cite{meibom2013}
provide an important test of how dynamical evolution shapes the architecture of planetary systems. 
\cite{meibom2013} find that the statistical population of planets in open clusters like NGC~6811 is indistinguishable from the field. 
More recent work has argued for the dynamical creation of hot Jupiters in star clusters \cite{shara2016} 
with some observational evidence for an increased incidence in high-density stellar clusters  \cite{brucalassi2016}. 
These studies provide an important link to the observed demographics of exoplanets, but may be dominated by later dynamical processing that shapes planetary architectures post-formation.

The majority of exoplanets are not found in clusters. 
There is a great deal of interest in inferring their birth environments. 
Such forensic studies are more difficult than for the Solar System where detailed meteoritics provide stringent constraints. 
Some efforts have been made to connect exoplanet properties with the impact of their current and past environments. 
A series of papers beginning with \cite{winter2020} use the local position-velocity phase-space density to distinguish between exoplanet hosts in `over-density' and `field' environments. 
This analysis finds statistically significant differences in the occurance rate of hot Jupiters in the field and over-density populations \cite{winter2020}. 
These statistical differences have also been used to argue that stellar clustering influences 
the planet radius valley \cite{kruijssen2020}, 
planetary multiplicity and orbital periods \cite{longmore2021}, and 
planet properties in multiple systems \cite{chevance2021}. 
These results have inspired a lively debate in the literature about whether age and metallicity bias the correlation  \cite{adibekyan2021,mustill2022}.

The present-day phase-space density does not necessarily reflect the birth environment. 
A recent contribution argues that the over-density regions are the result of large-scale (Galactic-scale) perturbations \cite{kruijssen2021}. 
Questions remain about the physical mechanism that drives the observed correlation. 
Densities in the `over-density' regions are much lower than in star-forming clusters, with a correspondingly low encounter rate (see discussion in \cite{kruijssen2021}). 
Interpreting the identified `over-density' and `field' regions in the context of known moving groups may help identify the relevant physical mechanism. 
Finally, a similar study investigating the demographics of stellar multiples in over-density and field regions would provide a larger statistical sample to investigate how phase-space overdensities affect the architectures of multiple systems.

\section{Future directions and future facilities}\label{s:future}

In the coming decade, large-scale surveys from multi-object spectroscopic facilities will provide observations of more stars in more environments. 
The new 4-metre Multi-Object Spectroscopic Telescope (4MOST) \cite{deJong2019} will be able to observe $\sim 2400$ objects over a 4.2$^{\circ}$ hexagonal field of view (FOV) and provide both moderate ($R \approx 4000-7500$) and high resolution ($R \approx 18,000-21,000$) spectra. 
Science goals for the project include measuring radial velocities with a precision of $<2$~\kms\ for \emph{Gaia} sources (${G_{\mathrm{Vega}}} < 20.5$~mag). 
Multiple community surveys with 4MOST
target young stars and star-forming regions to understand the formation, evolution, and dissolution of star clusters. 
4MOST is expected to start operations in early 2024. 

The Multi-object Optical and Near-IR spectrograph (MOONS; \cite{taylor2018}) will be installed on the 8~m Very Large Telescope (VLT) of the European Southern Observatory (ESO). 
MOONS makes use of the full 25$^{\prime}$ diameter FOV of the VLT, deploying 1001 fibres to obtain high resolution spectra ($R \approx 20,000$) with simultaneous optical and near-IR ($0.6-1.8$~$\mu$m) coverage. 
High-resolution coverage in the H-band will allow these observations to probe obscured regions, extending the work done at similar spectral resolution in nearby regions (e.g., by the IN-SYNC survey, \cite{cottaar2014,cottaar2015,foster2015,dario2016,dario2017}) to regions more distant and more massive than Orion. 
Probing earlier stages in the dynamical evolution of star clusters will help constrain models.

Full 6D measurements of fainter stars will be possible with additional data releases from the on-going \emph{Gaia} mission \cite{gaia_mission2016,gaia2016,gaia2018}. 
\emph{Gaia} will also continue to enable the discovery of previously unknown star clusters \cite{castro-ginard2021}. 
The current \emph{Gaia} optical mission has enabled us to assess the kinematics of exposed (gas-free) clusters and also quantified the numbers of runaway ($>30$~km\,s$^{-1}$) and walkaway ($>5$~km\,s$^{-1}$) stars on the outskirts of star-forming regions \cite{schoettler2020,schoettler2022}.  
In the future, a near-IR version of \emph{Gaia} would facilitate studies of the kinematics of star-forming regions at much earlier ages.
Targeted studies of star-forming regions within 500~pc from the 
VISTA Star Formation Atlas (VISIONS; \cite{meingast2016}) survey 
will provide multi-epoch near-IR imaging to enable proper motion measurements. 
Similar studies of more distant regions may be possible in the future with the wide field of view and sensitivity of the \emph{Roman Space Telescope}.

The recent launch of \emph{JWST} promises a wealth of near- and mid-IR observations of planet-forming disks. 
For Galactic high-mass star-forming regions beyond Orion, the angular resolution at these wavelengths will probably not suffice to spatially resolve proplyds, as discussed in Section~\ref{s:obs}. 
An additional complication for imaging studies is that the bright and variable nebulosity characteristic of these regions is likely to saturate in nearby regions. 
However, \emph{JWST} will enable the first detection of low-mass stars in more distant regions, including those with lower metallicity.  
Spectroscopy will be especially powerful for probing planet-forming disks affected by feedback, quantifying its impact on the disk structure, chemistry, and mineralogy (e.g., GO~1759, GO~1905).

Wide-field surveys of molecular gas with high angular resolution will be possible with future large single-dish telescopes like 
AtLAST \cite{klaassen2019_atlast}. 
The proposed 50~m facility will provide $\sim 5^{\prime\prime}$ angular resolution at 1~mm, a factor of $\sim 6$ better than the current best-available survey data. 
The corresponding spatial resolution at the distance of 
Taurus (140~pc) is $3 \times 10^{-3}$~pc or 700~AU; for
Orion (400~pc), 0.01~pc or 2000~AU; and for
Carina (2300~pc), 0.06~pc or 11,500~AU. 
Millimeter observations with angular resolution approaching that achieved at shorter wavelengths will 
enable more studies of the direct impact of feedback on the structure and kinematics of star-forming gas (e.g., \cite{goicoechea2016}). 
More comparable resolution for maps of the gas and stellar distributions in young clusters will also significantly improve inputs for simulations of their coevolution (e.g., \cite{sills2018}).

The COSI \cite{tomsick2022} mission, 
recently selected through NASA's astrophysics explorers program, will survey the sky in the soft $\gamma$-ray regime ($0.2-5$~MeV). 
COSI is a factor of $30-100$ more sensitive than the earlier generation of X-ray observatories   
with high spectral resolution to provide the the first map of \fer\ (1.173~MeV and 1.333~MeV) in the Galaxy and isolate \alr\ (1.809~MeV) from individual massive star clusters. 
The mission is currently in Phase~B development with an expected launch in late 2025.

Finally, multiple 30~m class telescopes (extremely large telescopes; ELTs) are under construction with first light expected at the end of the decade. 
The transformative sensitivity and angular resolution ELTs provide the first resolved observations of individual objects in regions too distant for high spectral and/or spatial resolution observations with current facilities. 
High-mass star-forming regions (with typical distances $\gtrsim 2$~kpc) will be prime targets for these facilities, providing direct tests of the impact of more intense feedback from multiple high-mass stars.

The improvement in angular resolution and sensitivity provided by upcoming facilities will unlock the study of low-mass stars in high-mass star-forming regions, sampling a broader range of feedback conditions. 
Many of these facilities are IR-optimized, opening up younger regions for detailed scrutiny. 
This is critical to determine how quickly feedback affects planet-forming disks. 
Finally, connecting the dots from feedback in the environment during planet formation and the observed demographics of exoplanets  will require creatively combining disparate observations.

For example, as discussed in Section~\ref{s:connections}, exoplanet host stars are nominally not traceable to a star-forming region, and thus we have very little information on the environment in which that host star was born. 
The recent phase-space study of exoplanet host stars by \cite{winter2020} postulated that the dynamically `hot' planetary systems correlated with dense phase space of the host star. 
While the interpretation that this is a relic of the birth environment is disputed \cite{kruijssen2021}, 
other planet properties may provide additional insight. 
A further novel way of constraining the birth environment of exoplanet hosts would be to accurately determine the radii of transiting terrestrial planets (e.g., with PLATO). 
Models with heating from short-lived radioactive isotopes (i.e., \alr)
predict that enriched systems have smaller radii \cite{lichtenberg2019}. 
For two planets of comparable composition, a planet without heating from  short-lived radioactive isotopes would have a larger radius as a result of its higher volatile content. 
If planets forming in high-mass star-forming regions are more likely to be enriched with \alr\ \cite{reiter2020_al}, this may provide another indirect indication of the birth environment.

For numerical simulations, environmental studies urgently need 
a critical appraisal of how effective gas/dust is at shielding disks from photoionizing radiation.
The high dust/gas content in star-forming regions may shield disks from evaporation, but this depends on how quickly the high-mass stars blow huge cavities -- as in the simulations of \cite{dale2011,dale2012} -- that rapidly leave the disks exposed to photoionizing radiation.  
Spatially-resolved observations of extinction in different clusters as a function of their age, size, and membership will be a key calibration of these models.
Another crucial piece of the puzzle will be 
whether we can accurately determine the true effects of feedback on protoplanetary disks. 
Photoevaporation affects both the gas  \cite{haworth2018_fried} and dust \cite{sellek2020} component of disks, but models do not yet include both effects.  
Extending models of external photoevaporation by, e.g.,  \cite{haworth2018_fried}, to maximum FUV fields $>10^4$~G$_0$ (the current limit) will also aid models of the time-evolution of star-forming regions like the ONC that likely had much higher initial FUV radiation fields \cite{parker2021_a}.

\section{Conclusions}\label{s:conclusions}

Studies of star-forming clusters as the birthplace of planets are experiencing a renaissance as more exoplanets are discovered and new facilities provide high-quality, large-scale survey data of star-forming regions. 
Most stars form in regions with higher densities than the Galactic field and will be affected by their stellar neighbors during the epoch of planet formation. 
Density is the strongest determinant of the impact of feedback. 
Dynamical interactions require high densities while external photoevaporation of planet-forming disks is important in both high- and low-density regions. 
How density evolves with time determines when and how long different feedback mechanisms affect disks, and thus disk lifetimes in the region as a whole.

Models are in broad agreement that feedback rapidly disperses disks. 
Observational confirmation is more challenging. 
Some studies show a clear spatial structure with fewer and/or smaller and lower mass disks near high-mass stars. 
Others do not, perhaps reflecting observational biases (projection effects), mixing of disk-bearing and disk-less sources by kinematics, and/or highly structured fields of radiation created by multiple high-mass stars.  
Recent work also suggests that star-forming regions may have spatially structured age differences, with younger sources concentrated toward the center. 
This may explain why disks are observed in regions with strong UV fields that should rapidly disperse them. 
If common, spatially structured age distributions will affect the apparent disk lifetime in feedback-dominated regions.  
While individual disks may be short-lived, disks may be present in the region for longer than the individual disk lifetime.

High-mass star-forming regions include thousands of low-mass stars. 
These are prime environments to quantify the impact of feedback but have historically been a challenge to observe given their typical distances of $\gtrsim 2$~kpc. 
As a result, there is growing interest in observational signatures of feedback on disks that does not require spatially-resolved images. 
Surveys of disk demographics in truly high-mass regions would be a helpful starting point, although to date, there have only been a few targeted studies of millimeter dust emission from disks in regions with high-mass stars. 
In the future, observations with ELTs may provide resolved observations to directly test a few sources.

Models of star-forming clusters now include multiple forms of feedback. 
These will provide better initial conditions for disk models. 
Higher resolution observations of the distribution of stars and gas in young regions will provide better inputs for N-body simulations coupled to a live background gas potential.

How frequently planet-forming systems are enriched with short-lived radioactive isotopes is of great interest given the important role these elements play in geochemical evolution. 
Galactic-scale observations reveal high abundances of elements like \alr\ in high-mass star-forming regions with dying high-mass stars as their likely source. 
Whether these elements are available early enough and mix efficiently with planet-forming material remains an open question. 
Recent observations of high-mass stars provide some additional constraints for enrichment models. 
For example, Galactic W--R stars (expected to be one of the primary sources of \alr) are infrequently found in star-forming regions. 
The observed rotation rates of high-mass stars are much slower than previously assumed in models, affecting when, how long, and from which stars \alr\ may be introduced into the local star-forming environment. 
These and other challenges have brought renewed interest to local \alr\ production by cosmic rays. 
High-mass stars themselves may be a significant source of cosmic rays and it would be interesting to consider this in models of \alr\ production and indeed their impact on star-forming regions more generally.

We highlight upcoming facilities that will provide much-needed observational constraints. 
Large-scale spectroscopic surveys will provide radial velocities, complementing the wealth of proper motion information from \emph{Gaia}. 
These data will also provide stellar parameters for thousands of young stars. 
New spectroscopic and imaging capabilities in the near-IR will be especially impactful as they will open up younger and more embedded regions, providing better constraints on the initial conditions in star-forming clusters.

\backmatter

\section*{Declarations}

\begin{itemize}
\item Funding:
MR acknowledges the support of an ESO fellowship. RJP acknowledges support from the Royal Society in the form of a Dorothy Hodgkin Fellowship.

\item Competing interests:
Not applicable.  

\item Ethics approval: 
Not applicable. 

\item Consent to participate:
Not applicable. 

\item Consent for publication:
Not applicable. 

\item Availability of data and materials:
Not applicable. 

\item Code availability: 
Not applicable. 

\item Authors' contributions:
Not applicable. 

\end{itemize}

\noindent
If any of the sections are not relevant to your manuscript, please include the heading and write `Not applicable' for that section.

\typeout{}
\bibliography{bibliography}


\end{document}